\documentclass[twocolumn,natbib]{svjour3}

\usepackage{graphicx}
\usepackage{amsmath}
\usepackage{amssymb}
\usepackage{ifpdf}\ifpdf 
\usepackage[colorlinks,linktocpage,pdftex,hyperfigures]{hyperref}
\makeatletter
\providecommand{\doi}[1]{%
  \begingroup
    \let\bibinfo\@secondoftwo
    \urlstyle{rm}%
    \href{http://dx.doi.org/#1}{%
      doi:\discretionary{}{}{}%
      \nolinkurl{#1}%
    }%
  \endgroup
}
\makeatother

\fi
\journalname{Microgravity Science and Technology}

\begin{document}

\title{Boiling crisis dynamics: low gravity experiments at high pressure}

\author{V. Nikolayev\and Y. Garrabos\and C. Lecoutre\and T. Charignon\and D. Hitz\and D. Chatain\and R. Guillaument\and S. Marre\and D. Beysens}
\authorrunning{V. Nikolayev \textit{et al.}} 

\institute{V. Nikolayev \at
              Service de Physique de l'Etat Condens\'e,\\ CNRS UMR 3680, IRAMIS/DSM/CEA Saclay,\\ 91191 Gif-sur-Yvette, France \\
               \email{vadim.nikolayev@cea.fr}
           \and
           Y. Garrabos \and C. Lecoutre\and R. Guillaument\and S. Marre \at
              CNRS, Univ. Bordeaux,\\
               ICMCB, UPR 9048,\\
               33600 Pessac, France\and
           T. Charignon \and D. Hitz\and D. Chatain\and D. Beysens\at
              Service des
Basses Temp\'eratures,\\ CEA-Universit\'e Grenoble Alpes, INAC,\\ 17 rue des Martyrs, 38054 Grenoble Cedex 9, France
\and D. Beysens\at Physique et M\'ecanique des Milieux H\'et\'erog\`enes,\\ UMR 7636 ESPCI - CNRS - Univ.~Paris-Diderot - Univ.~P.M.~Curie,\\ 10 rue Vauquelin, 75005 Paris, France}

\date{Received: \today / Accepted: }

\maketitle

\begin{abstract}
To understand the boiling crisis mechanism, one can take advantage of the slowing down of boiling at high pressures, in the close vicinity of the liquid-vapor critical point of the given fluid. To preserve conventional bubble geometry, such experiments need to be carried out in low gravity. We report here two kinds of saturated boiling experiments. First we discuss the spatial experiments with SF$_6$ at $46^\circ$C. Next we address two ground-based experiments under magnetic gravity compensation with H$_2$ at 33 K. We compare both kinds of experiments and show their complementarity. The dry spots under vapor bubbles are visualized by using transparent heaters made with metal oxide films. We evidence two regimes of the dry spots growth: the regime of circular dry spots and the regime of chain coalescence of dry spots that immediately precedes the heater dryout. A recent H$_2$ experiment is shown to bridge the gap between the near-critical and low pressure boiling experiments.
\end{abstract}


\section{Introduction}

When during boiling, the supplied heat flux exceeds a critical value (the Critical Heat Flux, CHF)
the vapor bubbles on the heating surface form abruptly a film that thermally insulates the heater from the
liquid. The heat transfer is blocked and the temperature of the heater rapidly grows. This transition is known
under the names of ``boiling crisis'' (BC) or ``departure from nucleate boiling''.

In the engineering literature, the term ``critical heat flux'' (CHF) is often used both as a synonym to the term ``boiling crisis'' (BC) and as a particular heat flux value. To avoid the ambiguity, we use the term CHF only to designate the threshold heat flux.

Correct CHF estimation requires a clear understanding of the physical
phenomena that triggers it. However, the number of parameters and interactions involved in a quantitative description of boiling is so large that no definitive theory has been achieved yet. Since the discovery of BC \citep{Nukiyama}, a large amount of both experimental data and theoretical models has been accumulated. On one hand, the mechanisms governing the bubble formation, growth and detachment, and their dependence on diverse parameters are quite clear \citep{Dhir}. On the other hand, the triggering factors of the vapor film formation are much more elusive.

During the last decade, advanced methods of observation and measurement (as high speed and high resolution optical and infrared cameras, microscopic thermal sensors and structured heating matrices, etc.) were applied to access experimentally the length scales smaller than the vapor bubbles and time scales below the bubble growth time.  A consensus about the local nature of BC has begun to emerge. Contrary to previous idea of BC as being triggered by bulk hydrodynamic phenomena \citep{Dhir}, the phenomena responsible for the BC are considered to act in a thin fluid layer adjacent to the heater \citep{Eulet99,Kandlikar01,Theofanous02,PRL06,Chung07,Gong14,Kim14,Yagov14,Kannengieser14}, down to the level of the triple (liquid-solid-vapor) contact. The accent is made on the growth dynamics of dry heater area. It is important to understand how drying impacts the BC triggering. Boiling crisis can be either triggered by the growth of dry spots under individual vapor
bubbles \citep{Eulet99,PRL06,IJHMT01,IPHT14} that coalesce later, or result from a collective phenomenon of the avalanche-like multiple bubble coalescences \citep{Lloveras12,PRE15}. It is still an open question that requires further investigations.

\section{Boiling crisis scenario}\label{BCS}

\begin{figure}[htb]
\centering
\includegraphics[width=0.6\columnwidth,clip]{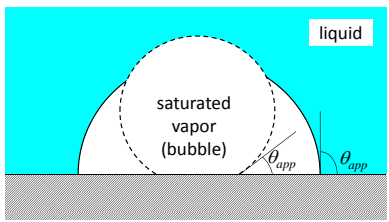}
\caption{Increase of the dry spot under a bubble caused by the increase of apparent contact angle.} \label{AppAngle}
\end{figure}
The dry spot formed on the heater under a vapor bubble depends on its shape defined by its apparent contact angle $\theta_{app}$ (Fig. \ref{AppAngle}). Under $\theta_{app}$ we mean the receding contact angle defined on a scale comparable to the bubble size. The relation between the apparent contact angle and the size of the dry spot was evidenced both experimentally \citep{Kandlikar} and by numerical simulation \citep{IJHMT01}. The apparent contact angle may differ from its equilibrium (usually small) value $\theta_{micro}$ \citep{PF10,PRE13}. It increases with the local heater temperature at the contact line position that varies with time. Therefore, $\theta_{app}$ increases during the bubble growth \citep{IPHT14}. At near critical pressure, where the hydrodynamic flow is very slow, the vapor recoil is the main cause of the apparent contact angle increase \citep{PRL06,PRE01}. On the contrary, farther away from the liquid-vapor critical point, the viscous pressure drop of the liquid flow toward the contact line is the main cause of the apparent contact angle increase \citep{PF10,PRE13}.

Based on the discussed above experimental observations, the BC scenario suggested by \cite{IPHT14} can be summarized here. At low heat fluxes, $\theta_{app}$ remains low throughout bubble growth; the bubble grows and departs from the heater under the gravity influence. At large heat flux, the apparent contact angle may attain $90^\circ$ during bubble growth. Such an event is important for the following reasons. First, the dry area attains its maximum value with respect to the bubble size leading to heater temperature rise. Second, the bubble adhesion to the heater increases, which facilitates bubble coalescence with neighboring bubbles. Third, and the most important, the dry areas under bubbles merge simultaneously with bubble coalescence, which leads to a fast heater temperature rise.

In the case $\theta_{app}<90^\circ$, the coalescence of neighboring bubbles does not result in immediate dry spot merging and spreading; the heater continues to be cooled under the newly formed bubble (because of the latent heat consumption) even after the coalescence of bubble volumes. The dry spot coalescence is slow because large viscous dissipation in the thin liquid film appearing in the contact line vicinity and pinning at solid surface defects causes the slowness of the contact line motion.

All these reasons suggest that the triggering of BC occurs when $\theta_{app}$ becomes larger than $90^\circ$ while the bubble grows on the heater. Therefore, CHF should correspond to the situation where $\theta_{app}$ attains $90^\circ$ exactly \emph{at} the moment of bubble departure. Such a criterion should hold for any boiling regime. The reason of such an universality is the smallness of the characteristic length scale (10-100 nm, see \citet{PRE13}) at which $\theta_{app}$ is formed. Therefore, the criterion is independent of other forces (e.g. gravity or inertial forces important at low pressure boiling) that are negligible at such small scales. Indeed, the theory of \cite{IPHT14} is corroborated by \cite{Kannengieser14} who observed $\theta_{app}\approx 90^\circ$ at CHF at low pressures. Note that a nearly equivalent criterion (the equality of the radii of curvature of the bubble and of the dry spot) has been adopted by \cite{Yagov14}.

Such a criterion allows the CHF value to be calculated as a function of various system parameters. A calculation that used a quasistatic approach in the bulk fluid (no fluid motion) allowed some of us calculating CHF as a function of wetting properties of the heater (i.e., $\theta_{micro}$) and gravity \citep{IPHT14}. As expected, such calculations led to zero CHF value at zero gravity in agreement of all previous approaches \citep{Dhir}. This means that under zero gravity, BC occurs whatever the supplied to fluid heat flux is, provided it does not vary in time and the experiment time is long enough.

Violent boiling conditions under which BC occurs complicate a detailed analysis and hinder a deeper understanding of the process. Experimental studies at normal conditions are very difficult to perform \citep{Theofanous02,Kim14} since the required heat fluxes are huge: CHF for water at atmospheric pressure is 1-10 MW/m$^2$. The main obstacle to observations is the highly non-equilibrium nature of the crisis, which makes impossible a steady state study. In addition, the BC characteristic time is very small, below 1 ms.

Near the critical point, thermal diffusivity tends to zero and  controlled by it bubble growth slows down. The overall liquid flow slows down either and the optical distortions associated with the strongly deformed vapor-liquid interfaces and turbulence disappear. In addition, the CHF becomes very small \citep{PRL06}, thus facilitating the experimental implementation. However, such boiling experiments are more complicated than those at low pressures. Since the surface tension reduces, the capillary length becomes small at normal gravity, which leads to flattening of the gas-liquid interfaces. Reduced gravity conditions are thus necessary to preserve conventional bubble shapes. While approaching closely the critical point, temperature should be stabilized with extremely high precision to avoid fluid transition to the single phase (i.e., supercritical) regime.

\section{DECLIC/ALI experiments}

The first series of experiments discussed here has been performed with the CNES DECLIC (French abbreviation for Dispositif pour l'Etude de la Croissance et des LIquides Critiques --
facility for studies of crystal growth and near-critical fluids) apparatus operated by NASA. It is located in the Kibo module of the International Space Station. It allows the use of different interchangeable modules (inserts). The ALI module (Alice-like insert) is called after the CNES Alice apparatus that functioned on board of the Mir space station \citep{PRE01}. ALI has been used for boiling experiments using pure SF$_6$ fluid, the critical temperature (i.e. that of the liquid-vapor critical point) of which is $T_c\simeq 46^\circ$C. ALI contains two sealed cells filled at nearly critical average density of SF$_6$, which means that near $T_c$ the gas and liquid volumes inside the cell are nearly equal independently of the cell temperature. One of the cells is the direct observation cell (DOC) where boiling can be observed through a transparent film heater \citep{ActaAstro10} supplying the power $P_h$. The second cell is the interferometry observation cell (IOC), where the light ray direction is, in contrast, parallel to the heater surface \citep{ActaAstro14}. Both cells are cylinders of $10.6$ mm inner diameter, with  transparent sapphire cylinder bases (except one of them, mirror reflective, in IOC) through which the cell interior can be observed. The cells are thermostated at the temperature $T<T_c$ within 10 $\mu$K precision, which means that the vapor generated with internal heaters recondense inside the cell. The ALI experiments have been successfully accomplished. They have produced a large array of video and other data which is currently under analysis.
\begin{figure}[htb]
\centering
\includegraphics[width=\columnwidth,clip]{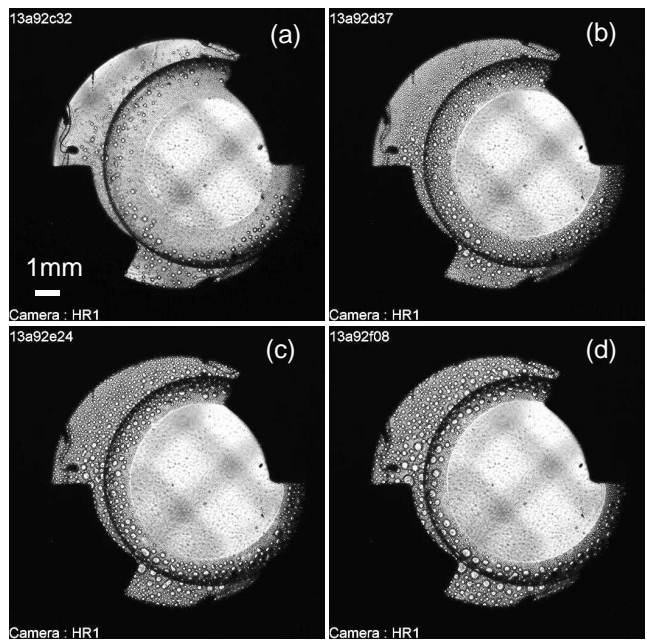}
\caption{Triple contact line receding observed in DECLIC (ALI/DOC). Observation is performed through the transparent metal oxide film heater, $P_h= 3$ mW and $T_c-T=20$ mK. The images correspond to about 2, 13, 23, and 33 s after the heating start for the images a,b,c,d, respectively. The hexadecimal time is shown at the upper left corner of each image using the internal DECLIC units $\approx 1/23$ s).} \label{ALI_SEQ2_3mW_20mK}
\end{figure}
We discuss here the DOC images. The thickness of DOC cylindrical cell (4.115 mm) is smaller than its diameter so the large vapor volume that would take, if unconstrained, the spherical shape, is squeezed between two transparent sapphire portholes. When the surface tension is strong enough (at $\Delta T\equiv T_c-T=20$ mK), the liquid meniscus is not deformed by the thermistors located inside and the visible meniscus shape (that we call ``large bubble'') is circular. At equilibrium, wetting is complete in this system: a continuous wetting film separates the bulk vapor from the heater. When heating begins, the wetting film evaporates completely at a part of the heater and a triple liquid-vapor-solid contact line appears. The heater drying dynamics is shown in Fig. \ref{ALI_SEQ2_3mW_20mK}. One can see nucleation, growth and coalescence of small bubbles at the heater. Small bubbles are absent in its central part (this could be checked with sufficient precision using the microscopy capability of DECLIC) indicating dry area.

At large $\Delta T$, the applied $P_h$ is not sufficient for the complete heater dryout. A major part of $P_h$ is absorbed by the temperature regulation because of too good thermal contact between the heater and the copper cell structure so that the heat flux supplied to the fluid decreases with time. The small bubbles remain circular, which indicates that each coalescence is an isolated event. In other words, the time between subsequent coalescences is large enough for complete bubble relaxation toward the spherical cap shape.

To observe complete dryout, one needs to increase the heating power or get closer to $T_c$.
\begin{figure}[htb]
\centering
\includegraphics[width=\columnwidth,clip]{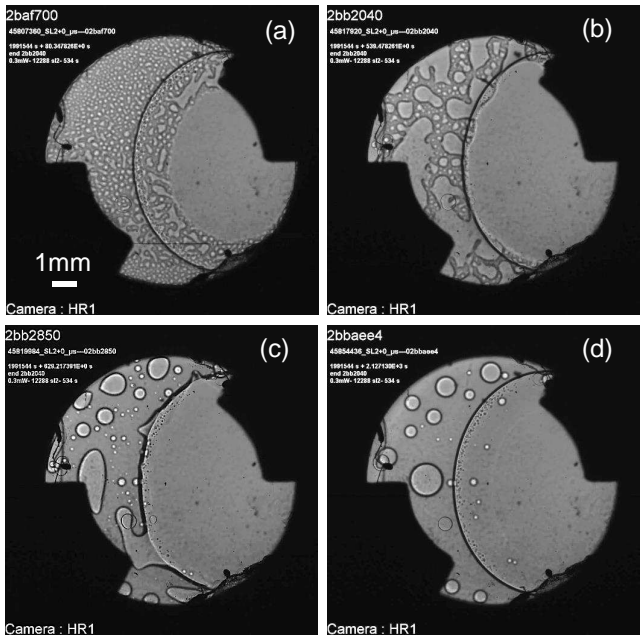}
\caption{Chain coalescence of dry spots at the heater observed in DECLIC for $P_h= 0.3$ mW and $T_c-T=0.6$ mK. The images correspond to about 75, 534, 624, and 2122 s after the heating start for the images a,b,c,d, respectively. The heating stops at the time moment corresponding to the image b.} \label{Ali9_Seq10_B24_0_3mW_0_6mK}
\end{figure}
The latter case is shown in Fig. \ref{Ali9_Seq10_B24_0_3mW_0_6mK}. Since the fluid is very close to the critical point, the evolution is extremely slow. First, one mentions that the apparent contact angle for all small bubbles is $\geq 90^\circ$ because the bubble contour and the contact line position coincide. This is coherent with the above BC scenario. Second, one observes a phenomenon of chain coalescence where the time between subsequent coalescences is smaller than the time of drop relaxation to the spherical cap shape. Such a process can lead to complete heater dryout if heating lasts long enough. The image shown in Fig. \ref{Ali9_Seq10_B24_0_3mW_0_6mK}d corresponds to the rewetting situation where the heater is recovered by the wetting film after heating has stopped; circular small vapor bubbles remain suspended inside the bulk liquid above the heater. One can distinguish them from the bubbles attached to the heater because of their rapid relaxation to the circular form, which is possible only when the contact line (that causes slow relaxation because of large viscous dissipation associated with its motion) disappears.

\begin{figure*}[htb]
\centering
\includegraphics[width=0.7\textwidth,clip]{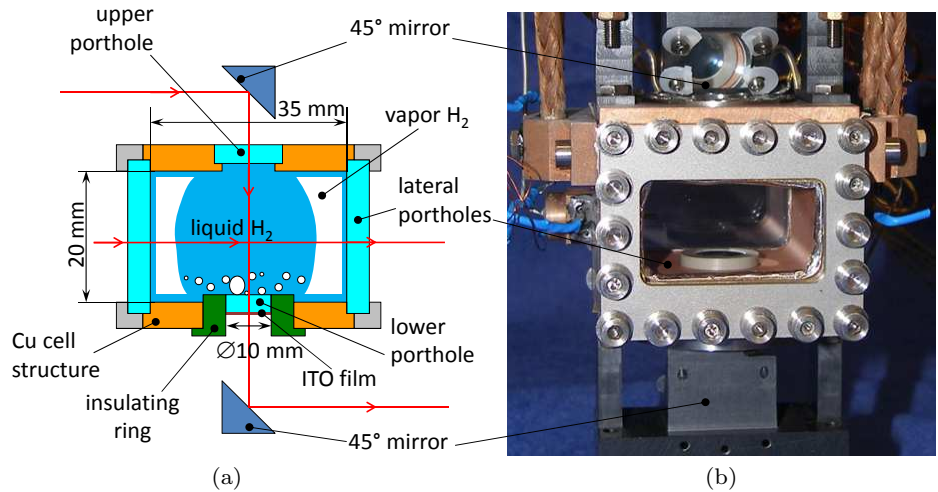}\\
(a)\hspace*{0.35\textwidth}(b)
\caption{(a) Sketch of the LHYLA cell with two way optical observation system. The gas-liquid distribution inside the cell in the critical point vicinity is also shown. (b) Cell photo. The sketch (a) corresponds to the cell section by a vertical plane perpendicular to the photo.} \label{figCompOpt}
\end{figure*}
We presume that a cross-over from the separate dry spot regime to chain coalescence always occurs during BC. It means that when the power is switched on, the separate dry spot regime occurs first which becomes later on (when the heater becomes hotter) the chain dry spot coalescence regime. When the fluid is far from the critical point, the duration of the latter regime is too small to be observed. When one approaches the critical point, a tiny heat flux is sufficient to immediately create the chain coalescence.

Further data treatment from DECLIC is needed to search such a cross-over.

\section{Magnetic gravity compensation results}

The DECLIC experiments provide valuable and accurate information on BC. However, because of the small volume limitation, the geometry and general boiling dynamics is quite different from  conventional pool boiling experiments at low pressure both at normal \citep{Chung07,Kim14} and  micro-gravity \citep{Straub01}. An experiment in a more conventional geometry (deep liquid pool) and a possibility of observation from the bottom and sidewise within the same experiment is needed to link the DECLIC and conventional boiling experiments. Ground based experiments with hydrogen have been carried out for this purpose. The gravity can be compensated in hydrogen by magnetic forces proportional to the square of the field gradient \citep{MST11}. The magnetic force per volume is also proportional to the fluid density so that the gravity can be compensated in the vapor and the liquid at the same time. It is however impossible to compensate the gravity uniformly in the whole volume. Typically, exact compensation occurs at only one point inside the cell. A special magnetic field configuration is chosen to minimize the residual gravity  in the whole volume. Its direction and space variation are defined by the magnetic field configuration. In all our experiments, the exact compensation point is chosen to situate in the heater center. The residual gravity is mainly centripetal; its vertical component is much smaller. The residual gravity strength with respect to the surface tension is characterized by the capillary length. On one hand, it is much larger than the small bubbles growing at the heater so that its influence on the boiling crisis development is negligible. On the other hand, near the critical point it is smaller than the cell size so that the residual gravity is dominant at this scale; liquid forms a column along the cell axis surrounded by gas.

\begin{figure}[htb]
\centering
\includegraphics[width=\columnwidth,clip]{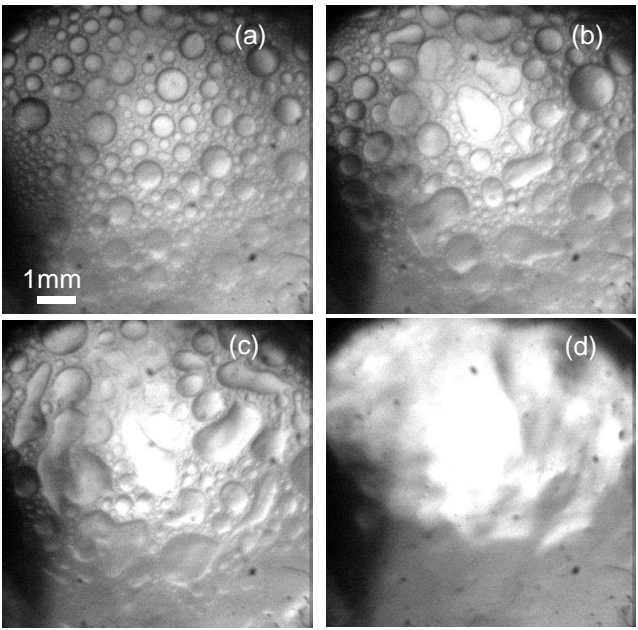}
\caption{Typical dry spot images recorded in one of the sequences of the HYLDE experiment  \citep{PRE15} at $T_c-T=50$ mK and 3.5 mW heater power. (a) Separate circular dry spots regime at 17.13 s before dryout. (b) Cross-over to the chain dry spot coalescence regime 7.2 s before dryout. (c) Chain coalescence regime 0.37 s before dryout. (d)  Dryout.} \label{HYLDE}
\end{figure}

We describe below a recent experiment performed with the LHYLA (Large HYdrogen Levitation Apparatus) installation that uses the M8 coil of the National laboratory of High Magnetic fields (LNCMI) at Grenoble. The experiment can be compared to the previous (2011) experiment carried out with the HYdrogen Levitation DEvice (HYLDE) described in detail by \cite{PRE15}.  The phase distribution in the LHYLA cell is sketched in Fig. \ref{figCompOpt}a. The experimental cell of parallelepiped shape is filled \emph{in situ} with pure hydrogen and then sealed. It has four sapphire portholes for the two-way optical observation (Fig. \ref{figCompOpt}a). Similarly to the DOC of DECLIC, the lower cylindrical porthole serves as a transparent heater because of the ITO film deposed on it from the bottom side. The heating surface is of the best possible quality, that of the bare sapphire (produced by the Rubis-Precis/Micropierre/HTC group; the surface roughness is 15 nm). It is thermally insulated from the copper cell structure with a PEEK (polyether ether ketone) ring to control the amount of heat injected into the fluid. Unlike the previous HYLDE experiment \citep{PRE15}, the cell is shaped in such a way that the heater is completely covered by the liquid, as in conventional boiling experiments. The liquid column bridges the top and the bottom portholes. Such a shape eliminates two problems of the HYLDE cell. The first is the partial heater dryout of the heater periphery well before the BC occurrence (the contact line formed by the liquid column edge is visible in the upper corners in Figs. \ref{HYLDE}), which prohibited the calculation of the actual heat flux supplied to the liquid. The second improvement is the elimination of the light refraction by a moving meniscus that has led to heterogeneous and time varying image brightness (Figs. \ref{HYLDE}) prohibiting the automatic image treatment. Indeed, there is no liquid meniscus in the optical path when observing through the transparent heater (Fig. \ref{figCompOpt}a). The cell is situated inside a cryostat under vacuum to thermally isolate the cell from the environment. The cell is surrounded by a screen cooled by the gaseous helium to prevent its radiative heating. Apart from the cell, the cryostat contains the hydrogen and helium circuits (the latter being used for cell  and screen cooling), the cell temperature regulation system and the displaceable periscopes for cell optical observation and lighting. The cryostat is centered inside the coil. The magnetic field used for gravity compensation is about 15 T. This coil provides  residual gravity less than $10^{-2}g$ within the useful for observations cylindrical volume of 10 mm diameter and 20 mm height. This level is three times better than in HYLDE.

The HYLDE results clearly showed the existence of two regimes of dry spot growth at the heater: the chain dry spot coalescence regime (Fig. \ref{HYLDE}c) similar to Figs. \ref{Ali9_Seq10_B24_0_3mW_0_6mK}a,b, and the separate circular spots regime (Fig. \ref{HYLDE}a similar to Figs. \ref{ALI_SEQ2_3mW_20mK}c,d). Similarity with the DECLIC experiments is striking. However, in the HYLDE experiments, a crossover between two regimes has been observed (Fig. \ref{HYLDE}b). In other words, the circular spots, the chain dry spot coalescence and the transition betwwen them have been observed on one occasion within the same time sequence (Figs. \ref{HYLDE}). This suggests that the chain dry spot coalescence always precedes the dryout; farther from the critical pressure, the chain dry spot coalescence regime becomes however too short to be observed. Close to the critical point, the circular spots regime becomes too short.

\begin{figure}[htb]
\centering
\includegraphics[width=\columnwidth,clip]{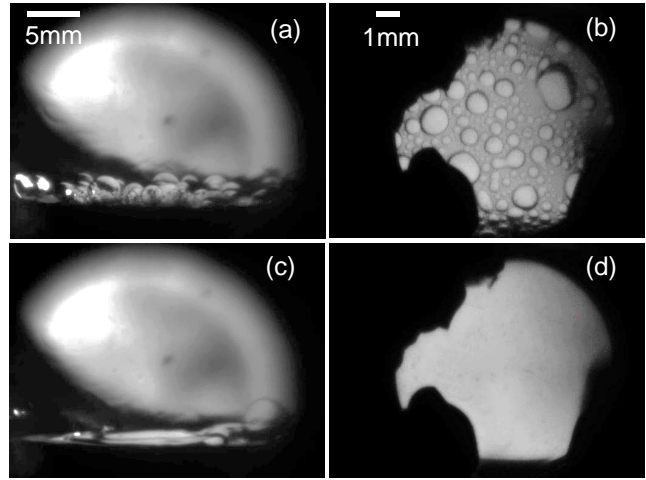}
\caption{Boiling of hydrogen under magnetic gravity compensation: the results of the 2014 LHYLA campaign for $T_c-T=80$ mK and $P_h= 44$ mW. Corresponding images obtained from sidewise observations (a,c) and through the transparent heater (b,d) just before BC in the nucleate boiling regime (a,b) and after BC in the film boiling regime (c,d). The heater situates at the bottom of the photos a,c.} \label{LNCMI14_80mK}
\end{figure}

The first campaign of LHYLA measurements has been carried out in October 2014. The circular spots regime (Figs. \ref{LNCMI14_80mK}b,d) has been obtained. When the lower porthole is heated, bubbles nucleate and grow on it. Rather than vertical bubble departure, residual gravity causes horizontal displacement of bubbles toward the lateral surface of liquid column. The bubbles merge there with the bulk vapor. The vapor recondenses at the surface of the wetting film that covers the internal surface of the copper cell structure. Its temperature $T$ is regulated to be constant thus imposing the pressure inside the cell to be the saturation pressure for this temperature, just as in DECLIC. For the first time we could observe the film boiling at the microgravity conditions (Figs. \ref{LNCMI14_80mK}c,d). The 2014 campaign showed the feasibility of such experiments. In the next stage, the thermal regulation system needs to be improved to approach closer the critical point where the chain dry spot coalescence regime is expected.

One notices that the dry spot behavior observed in the LHYLA experiments (Figs. \ref{LNCMI14_80mK}b,d) is completely analogous to both the HYLDE and DECLIC experiments described above. At the same time, the sidewise observations (Figs. \ref{LNCMI14_80mK}a,c) are similar to conventional (low pressure) boiling observations in microgravity \citep{Straub01}. The bubbles nucleate and grow on the heater but do not detach from it. However, unlike the spatial low pressure experiments, no hovering above the heater vapor mass is formed. Its absence is explained by the centripetal residual gravity forces that cause the nucleated vapor bubbles to move toward the cell periphery (cf. Fig. \ref{figCompOpt}a; the vapor mass is out of the field of view in Figs. \ref{LNCMI14_80mK}a,c). Note that the effective gravity level may be varied by changing the magnetic field. When the effective gravity increases, the bubbles depart vertically from the heater just like in the conventional boiling experiments.

\section{Conclusions}

Experiments near the critical point present a powerful and versatile tool. They reveal new details of boiling at high heat flux (i.e. near CHF), otherwise difficult to observe. Our observations are incompatible with theories of the boiling crisis relying on bulk or bubble hydrodynamics, in particular with the Zuber model and macrolayer evaporation theories developed by Katto et al. \citep{Dhir}. Indeed we don't observe vapor stems or even macrolayer (numerous bubbles separating the heater from a large hovering bubble). However the heater dryout occurs inevitably. The triggering phenomenon of the boiling crisis is thus related to the dry spot spreading; the relevant mechanism is microscopic and should act in the vicinity of the triple vapor-liquid-heater contact line. Just before the boiling crisis, the apparent contact angle is close to $90^\circ$. This observation is coherent with the proposed earlier CHF criterion ($90^\circ$ apparent contact angle at the moment of bubble departure).

Two regimes of dry spot growth are identified: circular dry spot growth observed far from the boiling crisis) and chain dry spot coalescence regime that (as we assume) always precedes the boiling crisis. The objective of the future studies is to study the crossover between these regimes.

While the data obtained in the magnetic gravity compensation experiments are much less precise than in the spatial experiments, they are much cheaper. Due to their versatility, magnetic gravity compensation experiments are important to bridge the gap between spatial near-critical and conventional boiling experiments.

\section*{Acknowledgments}

The financial support of CNES within the fundamental microgravity research program and two EMFL grants that covered usage of the LNCMI magnet facility are acknowledged. The authors are grateful to J. Chartier and P. Bonnay of SBT for the help with the development and technical support of the magnetic gravity compensation experiments. We acknowledge the support of the LNCMI, member of the European Magnetic Field Laboratory (EMFL). We thank the whole CNES-DECLIC team and in particular G. Pont for their enthusiastic and helpful involvement in this work.


\end{document}